%% file: 201511PRAcommentSuzuki06arxiv.tex
\documentclass[%
 reprint,
 amsmath,amssymb,
 aps,
prl,
showpacs
]{revtex4-1}

\usepackage{graphicx}
\usepackage{dcolumn}
\usepackage{bm}
\usepackage[bookmarkstype=toc,colorlinks=true,urlcolor=blue,linkcolor=blue,citecolor=blue,linktocpage=true,bookmarks=true,setpagesize=false]{hyperref}

\usepackage{braket}

\def\rmi{{\rm i}}

\begin{document}

\preprint{APS/123-QED}

\title{Analysis of Pancharatnam--Berry phase of vector vortex states \\via Pancharatnam connection}
\thanks{A footnote to the article title}%





\author{Masato Suzuki}
\affiliation{Department of Applied Physics, Hokkaido University, Kita-13, Nishi-8, Kita-ku, Sapporo 060-8628, Japan}
\author{Keisaku Yamane}
\affiliation{Department of Applied Physics, Hokkaido University, Kita-13, Nishi-8, Kita-ku, Sapporo 060-8628, Japan}
\affiliation{JST, CREST, Kita-13, Nishi-8, Kita-ku, Sapporo 060-8628, Japan}
\author{Kazuhiko Oka}
\affiliation{Department of Applied Physics, Hokkaido University, Kita-13, Nishi-8, Kita-ku, Sapporo 060-8628, Japan}
\author{Yasunori Toda}
\affiliation{Department of Applied Physics, Hokkaido University, Kita-13, Nishi-8, Kita-ku, Sapporo 060-8628, Japan}
\affiliation{JST, CREST, Kita-13, Nishi-8, Kita-ku, Sapporo 060-8628, Japan}
\author{Ryuji Morita}
\email{morita@eng.hokudai.ac.jp}
\affiliation{Department of Applied Physics, Hokkaido University, Kita-13, Nishi-8, Kita-ku, Sapporo 060-8628, Japan}
\affiliation{JST, CREST, Kita-13, Nishi-8, Kita-ku, Sapporo 060-8628, Japan}

\date{\today}

\textbf{Comment on ``Higher Order Pancharatnam-Berry Phase and the Angular Momentum of Light''}

In a recent letter, Milione \textit{et al.} reported the Pancharatnam--Berry phase (PBP) on a higher-order Poincar\'e sphere (HOPS). They used two \textit{spin--orbit} converters (SOCs) \cite{PhysRevE.60.7497} to make a trajectory along the HOPS. However, the SOCs actually do not make a path on the HOPS. We briefly discuss a way of acquiring a PBP made by SOCs.

\begin{figure}[b]
\includegraphics[width=0.45\textwidth]{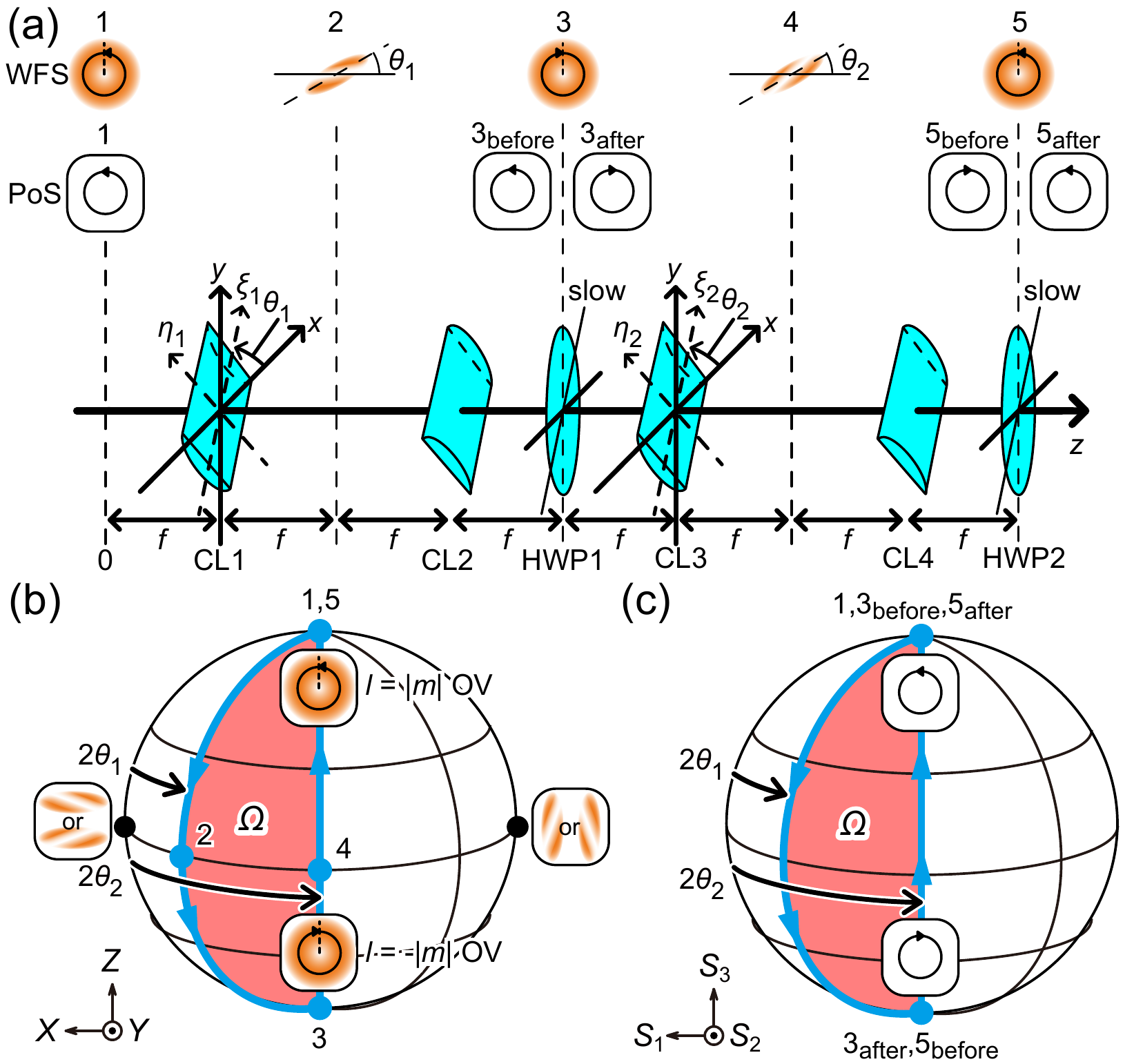}
	\caption{\label{fig:setup} (a) Configuration of the \textit{spin--orbit} converters. CL1,2,3,4: cylindrical lenses; HWP1,2: half-wave plates. WFS and PoS are abbreviations for the wavefunction state and polarization state, respectively. (b) Sphere for a $\pi$ converter and  (c) the conventional Poincar\'e sphere. This figure is drawn for $\sigma=1$. }
\end{figure}

Figure~\ref{fig:setup}(a) shows the configuration of the two SOCs. The first and second SOCs are rotated around the propagation ($z$) axis by $\theta_1$ and $\theta_2$ with respect to the $x$ axis, respectively. We define the rotating frame as
\begin{equation}
	\begin{pmatrix}\xi_i&\eta_i\end{pmatrix}^\mathrm{T} = R_{-\theta_i}\begin{pmatrix}x_i&y_i\end{pmatrix}^\mathrm{T}\quad(i=1,2),
\end{equation}
where $R_\theta$ is a rotation matrix.
In this frame, the reflection axis of the 4$f$-cylindrical lens pair ($\pi$ converter) and the slow axis of the half-wave plate (HWP) in the $i$th SOC are chosen to be the $\xi_i$ axis.

Here, we can divide the PBP into two PBPs made by $\pi$ converters and by two HWPs. First, we discuss the former one. A $\pi$ converter transforms the coordinate system as follow:
\begin{equation}
	\begin{pmatrix}x'\\y'\end{pmatrix} = R_{\theta_i}\begin{pmatrix}1&0\\0&-1\end{pmatrix}R_{-\theta_i} \begin{pmatrix}x\\y\end{pmatrix}\quad(i=1,2).
\end{equation}
We assume that the incident beam is collimated and written as $g(x,y,z)\begin{pmatrix}1&\sigma\rmi\end{pmatrix}^\mathrm{T}\propto e^{\rmi l\phi}\begin{pmatrix}1&\sigma\rmi\end{pmatrix}^\mathrm{T}$ at $z=0$. The wavefunctions $g$ at $z=4f$ and $8f$ are respectively expressed as
\begin{align}
	g(x,y,4f) &= g(x',y',0) \propto e^{2\rmi l\theta_1}g(x,y,0),\\
	g(x,y,8f) &= g(x',y',4f) \propto e^{-2\rmi l\theta_2}g(x,y,4f)\nonumber \\
&\quad\propto e^{2\rmi l(\theta_1-\theta_2)}g(x,y,0).
\end{align} 
Thus, the PBP made by the $\pi$ converters is $2 l(\theta_1-\theta_2) = -\Omega_\text{f}/2$, where $\Omega_\text{f}$ is subtended by the trajectory on the sphere for the $\pi$ converter \cite{Padgett:99}.

The intermediate wavefunction states in the $\pi$ converters at point 2 ($z=2f$) and point 4 ($z=6f$), which are discussed in Ref.~\cite{Denisenko:09}, are far from vector vortex sates. Thus, the intermediate states in the $\pi$ converters are not described on the higher-order Poincar\'e sphere but the sphere for a $\pi$ converter (Fig.~\ref{fig:setup}(b)). Although the states and the PBP on or along the sphere for $\pi$ converters are not well analyzed yet, we think that the PBP is related to the PBP made by the Gouy phase shift \cite{PhysRevLett.70.880} and by perfect reflection \cite{berry1987interpreting,*PhysRevLett.58.523}.

Needless to say, the latter PBP, which is made by two HWPs, is given by $-\Omega_\text{l}/2$, where $\Omega_\text{l}$ is the area enclosed by the trajectory on the conventional Poincar\'e sphere (Fig.~\ref{fig:setup} (c)). By comparing Fig.~\ref{fig:setup}(b) with Fig.~\ref{fig:setup}(c), we found that $\Omega_\text{f} = \Omega_\text{l} (\equiv\Omega)$. Therefore, we obtain the total PBP as
\begin{equation}
	\gamma(C) =  -l\Omega_\text{f}/2 - \sigma\Omega_\text{l}/2 = -(l+\sigma)\Omega/2,
\end{equation}
which agrees with the experimental results in the letter we comment on. We will report on the PBP made by $q$-plates, whose trajectories are truly along the HOPS elsewhere.

\noindent Masato Suzuki, Keisaku Yamane, Kazuhiko Oka, Yasunori Toda, and Ryuji Morita\\
\indent Department of Applied Physics\\
\indent Hokkaido University\\
\indent Kita-13, Nishi-8, Kita-ku, Sapporo 060-8628, Japan


\input{201511PRAcommentSuzuki06arxiv.bbl}

\end{document}

%% file: 201511PRAcommentSuzuki06arxiv.bbl
\providecommand{\noopsort}[1]{}\providecommand{\singleletter}[1]{#1}%
%